\documentclass[reqno,12pt]{amsart}

\def\be{\begin{equation}}
\newcommand{\dd}{{\rm d}}
\newcommand{\e}[1]{\,{\rm e}^{#1}\,}
\newcommand{\ii}{{\rm i}}
\newcommand{\abs}[1]{\lvert #1 \rvert}

\def\Tr{{\operatorname{Tr\,}}}
\def\bra #1{\langle#1 |\,}
\def\ket #1{\,|#1 \rangle}
\newcommand{\expval}[1]{\langle #1 \rangle}
\newcommand{\sumtwo}[2]{\sum_{\substack{#1 \\ #2}}}
\newcommand{\caD}{{\mathcal D}}
\newcommand{\bsS}{{\boldsymbol S}}
\newcommand{\bbZ}{{\mathbb Z}}
\begin{document}

\begin{quote}
\raggedleft
\end{quote}
\vspace{2mm}

\title[Coexistence of long-range order for two observables ...]
{Coexistence of long-range order for two observables at finite temperatures}

\author[N. Macris and C.-A. Piguet]{N. Macris and C.-A. Piguet}

\maketitle

\begin{centering}
{\it Institut de Physique Theorique \\
Ecole Polytechnique F\'ed\'erale de Lausanne \\
PHB-Ecublens, CH-1015 Lausanne, Switzerland\\}
\end{centering}

\thispagestyle{empty}

\vspace{10mm}

\noindent{\bf Abstract}

We give a criterion for the simultaneous existence or non existence of two
long-range orders for two observables, at finite temperatures, for quantum
lattice many body systems. Our analysis extends previous results of G.-S. Tian
limited to the ground state of similar models. The proof involves an inequality
of Dyson-Lieb-Simon which connects the Duhamel two-point function to the usual
correlation function. An application to the special case of the Holstein model
is discussed.

\vspace{1mm}

\newpage

\setcounter{page}{1}

\section{Introduction}

In a recent paper, G.-S. Tian derived a sufficient condition for the
simultaneous existence or non existence of long-range order of two observables
in the ground state of quantum lattice  models \cite{Tian1}. Here we prove a
similar criterion for the case of finite temperatures. As will be seen the
theorem is quite general, but for concreteness we limit ourselves to a specific
class of fermion Hamiltonians given by (\ref{hamiltonian}) below. A concrete
example concerning the Holstein model is given at the end and an application to
the comparison of critical temperatures for systems with more than one order
parameter is briefly discussed.

Let us first recall the result of \cite{Tian1}. Let $H_{\Lambda}$ be a fermion
Hamiltonian of the form
\be
H_{\Lambda}=\sumtwo{x,y\in\Lambda}{\sigma=\uparrow,\downarrow}t_{xy}c_{x\sigma}^{+}
c_{y\sigma}+\sum_{x,y\in\Lambda}V_{xy}n_{x}n_{y}+\sum_{x,y\in\Lambda}J_{xy}\bsS_{x}
\cdot\bsS_{y}
\label{hamiltonian}
\end{equation}
where $c_{x\sigma}^{+}$, $c_{y\sigma}$ are creation and annihiliation operators
of spin one-half fermions on a $d$-dimensional lattice $\Lambda\in\bbZ^{d}$,
$n_{x}$ and $\bsS_x$ are the usual fermion density and spin at site $x$. We
will always assume that the hopping matrix $t_{xy}$ and the two body
interactions $V_{xy}$, $J_{xy}$ are integrable. This implies that 
$\frac{1}{\abs{\Lambda}}\sum_{x,y\in \Lambda}\abs{a_{xy}}=O(1)$ with respect
to $\abs{\Lambda}$, where $a_{xy}=t_{xy},V_{xy},J_{xy}$.

Let $\ket{\Psi_{0}^{\Lambda}}$ be the ground state of $H_{\Lambda}$. If $O_x$
is a local operator, i.e. an operator valued function of
$\{c_{y\sigma}^{\#},y\in\caD_x\}$ for $\caD_x$ a bounded neighborhood of
$x$ such that $\abs{ \caD_x}=O(1)$ with respect to the volume $\abs{\Lambda}$,
we define the Fourier transform for each vector $k$ in the reciprocal lattice as
\be
\hat{O}_{k}=\frac{1}{\sqrt{\abs{\Lambda}}}\sum_{x\in\Lambda}\e{-\ii k\cdot x}O_x
\end{equation}
Suppose there exist three local observables  $A_x$, $B_x$, $C_x$ such that for
each $x\in\Lambda$
\be
[H_{\Lambda},A_x]=\mu B_x+\nu C_x
\label{commutation}
\end{equation}
where $\mu$ and $\nu$ are two non-vanishing complex numbers. The result of
\cite{Tian1} states that if (\ref{commutation}) and some additional assumptions
(see later on) are satisfied then
\be
\lim_{\Lambda\rightarrow\bbZ^{d}}\frac{1}{\abs{\Lambda}}\bra{\Psi_{0}^{\Lambda}}
\hat{B}_{k}^{+}\hat{B}_{k}\ket{\Psi_{0}^{\Lambda}}\neq 0
\label{Border}
\end{equation}
if and only if
\be
\lim_{\Lambda\rightarrow\bbZ^{d}}\frac{1}{\abs{\Lambda}}\bra{\Psi_{0}^{\Lambda}}
\hat{C}_{k}^{+}\hat{C}_{k}\ket{\Psi_{0}^{\Lambda}}\neq 0
\label{Corder}
\end{equation}
Notice that if (\ref{Border}) and (\ref{Corder}) do not vanish they are
strictly positive.
In fact (\ref{Border}) and (\ref{Corder}) express the presence of
long-range order for the observables $B_x$ and $C_x$. Indeed
\be
\frac{1}{\abs{\Lambda}}\bra{\Psi_{0}^{\Lambda}}\hat{O}_{k}^{+}\hat{O}_{k}
\ket{\Psi_{0}^{\Lambda}}=
\frac{1}{\abs{\Lambda}^2}\sum_{x,y\in\Lambda}\e{\ii k\cdot(y-x)}
\bra{\Psi_{0}^{\Lambda}}O_{y}^{+}O_{x}\ket{\Psi_{0}^{\Lambda}}\leq
\frac{1}{\abs{\Lambda}^2}\sum_{x,y\in\Lambda}\abs{\bra{\Psi_{0}^{\Lambda}}
O_{y}^{+}O_{x}\ket{\Psi_{0}^{\Lambda}}}
\label{6}
\end{equation}
Therefore if the left-hand side of (\ref{6}) is strictly positive one must also
have
\be
\lim_{\abs{x-y}\rightarrow\infty}\lim_{\Lambda\rightarrow\bbZ^{d}}
\abs{\bra{\Psi_{0}^{\Lambda}}O_{y}^{+}O_{x}\ket{\Psi_{0}^{\Lambda}}}\geq\epsilon>0
\end{equation}
for some $\epsilon>0$.

The result of \cite{Tian1} is based on the following inequality which follows
from the commutator relation (\ref{commutation}). For any $k$
\begin{multline}
\left(\abs{\mu}\sqrt{
\bra{\Psi_{0}^{\Lambda}}\hat{B}_{k}^{+}\hat{B}_{k}\ket{\Psi_{0}^{\Lambda}}}-
\abs{\nu}\sqrt{\bra{\Psi_{0}^{\Lambda}}\hat{C}_{k}^{+}\hat{C}_{k}\ket{\Psi_{0}^{\Lambda}}}\right)^2
\leq\bra{\Psi_{0}^{\Lambda}}[\hat{A}_{k}^{+},\hat{K}_{k}]\ket{\Psi_{0}^{\Lambda}}^{1/2}\\
\cdot\bra{\Psi_{0}^{\Lambda}}[\hat{K}_{k}^{+},[H,\hat{K}_{k}]]
\ket{\Psi_{0}^{\Lambda}}^{1/2}
\label{tianineq}
\end{multline}
where $\hat{K}_{k}=[H_{\Lambda},\hat{A}_{k}]$. If the right hand side of (\ref{tianineq})
is $O(1)$ uniformly in $\abs{\Lambda}$, the two terms in the square are
necessarily of the same order.  More precisely assuming that the right-hand
side is $O(1)$, if one of the correlation
functions is $O(\abs{\Lambda})$, this is also the case for the other. The
reader is referred to \cite{Tian2} for several concrete examples. Let us note
that if $\nu$ vanishes (resp. $\mu$ vanishes) (\ref{tianineq}) implies that
$\bra{\Psi_{0}^{\Lambda}}\hat{B}_{k}^{+}\hat{B}_{k}\ket{\Psi_{0}^{\Lambda}}=O(1)$ for all
$k$ which means there is no long-range order for $B_x$ (resp. $C_x$)
\cite{Tian3}. In the next section we prove a similar criterion for the case of
finite temperatures.

\section{Finite temperatures}

Let $\expval{O_{x}}_{\Lambda}=\Tr(O_{x}\e{-\beta(H_{\Lambda}-\mu
N_{\Lambda})})/Z_{\Lambda}$ where the trace is over the usual Fock space,
$N_{\Lambda}=\sum_{x\in\Lambda}n_x$ and $Z_{\Lambda}=\Tr\e{-\beta H}$ is the
partition function at inverse temperatures $\beta$.
The generalisation of (\ref{tianineq}) to
finite temperature involves the Duhamel two-point function whose definition we
recall here.
The Duhamel two-point function of two operators $F$ and $G$ is defined by
\be
(F,G)_{\Lambda}=\frac{1}{Z_{\Lambda}}\int_{0}^{1}\dd v
\Tr{\left(\e{-v\beta (H_{\Lambda}-\mu N_{\Lambda})}F\e{-(1-v)\beta(H_{\Lambda}
-\mu N_{\Lambda})}G\right)}
\end{equation}
It is symmetric $(F,G)_{\Lambda}= (G,F)_{\Lambda}$
and satisfies a Cauchy-Schwarz inequality
\be
\abs{(F,G)_{\Lambda}}^{2}\leq(F^{+},F)_{\Lambda}(G^{+},G)_{\Lambda}
 \end{equation}
We also recall the following useful identity
\be
\expval{[F,G]}_{\Lambda}=([F,\beta H_{\Lambda}],G)_{\Lambda}
\label{th-du1}
\end{equation}
The following lemma generalizes the inequality
(\ref{tianineq}).

\vspace{5mm}

{\bf Lemma}

{\it Let $A_x$, $B_x$ and $C_x$ be three local observables satisfying
(\ref{commutation}) then for all $k$
\be
\left(\abs{\mu}\sqrt{(\hat{B}^{+}_{k},\hat{B}_{k}})_{\Lambda}-\abs{\nu}
\sqrt{(\hat{C}^{+}_{k},\hat{C}_{k}})_{\Lambda}\right)^{2}
\leq\frac{1}{\beta}\expval{[\hat{A}^{+}_{k},\hat{K}_{k}]}_{\Lambda}
\label{ineqduhamel}
\end{equation}}

{\bf Proof:}

Setting $F=\hat{A}_{k}^{+}$ and $G=[H_{\Lambda},\hat{A}_k]=\hat{K}_k$ in (\ref{th-du1})
\be
\frac{1}{\beta}\expval{[\hat{A}^{+}_{k},[H_{\Lambda},\hat{A}_{k}]]}_{\Lambda}=
([\hat{A}^{+}_{k},H_{\Lambda}],[H_{\Lambda},\hat{A}_{k}])_{\Lambda}=
([H_{\Lambda},\hat{A}_{k}]^{+},[H_{\Lambda},\hat{A}_{k}])_{\Lambda}
\label{th-du2}
\end{equation}
Replacing the Fourier transform of (\ref{commutation}) in the right-hand side
of (\ref{th-du2}) we obtain
\begin{multline}
\frac{1}{\beta}\expval{[\hat{A}^{+}_{k},\hat{K}_{k}]}_{\Lambda}
=(\mu^{*} \hat{B}^{+}_{k}+\nu^{*}\hat{C}^{+}_{k},\mu \hat{B}_{k}
+\nu \hat{C}_{k})_{\Lambda}=\\
\abs{\mu}^{2}(\hat{B}_{k}^{+},\hat{B}_{k})_{\Lambda}+
\abs{\nu}^{2}(\hat{C}^{+}_{k},\hat{C}_{k})_{\Lambda}
+\mu^{*}\nu(\hat{B}_{k}^{+},\hat{C}_{k})_{\Lambda}
+\nu^{*}\mu(\hat{C}_{k}^{+},\hat{B}_{k})_{\Lambda}
\label{ineq1}
\end{multline}
We can estimate the last two terms on the right-hand 
side of (\ref{ineq1}) using
the Cauchy-Schwarz inequality
\begin{multline}
\abs{\mu^{*}\nu(\hat{B}_{k}^{+},\hat{C}_{k})_{\Lambda}+\nu^{*}
\mu(\hat{C}_{k}^{+},\hat{B}_{k})_{\Lambda}}
\leq\abs{\mu\nu}(\abs{(\hat{B}^{+}_{k},\hat{C}_{k})_{\Lambda}}
+\abs{(\hat{C}^{+}_{k},\hat{B}_{k})_{\Lambda}})\\
\leq 2\abs{\mu\nu}\sqrt{(\hat{B}^{+}_{k},\hat{B}_{k})_{\Lambda}}
\sqrt{(\hat{C}^{+}_{k},\hat{C}_{k})_{\Lambda}}
\end{multline}
Noticing that the term $\mu^{*}\nu(\hat{B}^{+}_{k},\hat{C}_{k})_{\Lambda}
+\nu^{*}\mu(\hat{C}^{+}_{k},\hat{B}_{k})_{\Lambda}$ is in fact
real, we have
\be
\mu^{*}\nu(\hat{B}_{k}^{+},\hat{C}_{k})_{\Lambda}+\nu^{*}\mu(\hat{C}_{k}^{+},
\hat{B}_{k})_{\Lambda}
\geq -2\abs{\mu\nu}\sqrt{(\hat{B}_{k}^{+},\hat{B}_{k})_{\Lambda}}
\sqrt{(\hat{C}_{k}^{+},\hat{C}_{k})_{\Lambda}}
\label{ineq2}
\end{equation}
From (\ref{ineq1}) and  (\ref{ineq2}) we get
\be
\frac{1}{\beta}\expval{[\hat{A}^{+}_{k},\hat{K}_{k}]}_{\Lambda}
\geq \left(\abs{\mu}\sqrt{(\hat{B}^{+}_{k},\hat{B}_{k})_{\Lambda}}
-\abs{\nu}
\sqrt{(\hat{C}_{k}^{+},\hat{C}_{k})_{\Lambda}}\right)^{2}
\end{equation}
which is the desired inequality.
To extract information about the correlation functions of $B_x$ and $C_x$ we
have to connect these to the Duhamel two-point function. This will be done
through
upper and lower bounds on the Duhamel two-point
function which involves only the symmetrized correlation function because the
Duhamel two-point function is symmetric. The upper bound follows from
convexity
\be
(F^{+},F)_{\Lambda}\leq\frac{1}{2}\expval{F^{+}F+FF^{+}}_{\Lambda}
\label{upperbound}
\end{equation}
The lower bound was found by Dyson, Lieb and Simon \cite{DLS}
\be
(F^{+},F)_{\Lambda}\geq\frac{1}{2}\expval{F^{+}F+FF^{+}}_{\Lambda}
f(h_{\Lambda}(F))
\label{lowerbound}
\end{equation}
where
\be
h_{\Lambda}(F)=
\frac{\expval{[F^{+},[\beta H_{\Lambda},F]]}_{\Lambda}}
{2\expval{F^{+}F+FF^{+}}_{\Lambda}}=
\end{equation}
and the function $f(u)$ is defined implicitely for $u>0$ by
\be
f(u\tanh{u})=\frac{1}{u}\tanh{u}
\end{equation}
The function $f(u)$ is continuous convex and strictly decreasing with
$\lim_{u\rightarrow 0}f(u)=1$ and $\lim_{u\rightarrow\infty}f(u)=0$.

For any local observable $O_x$, we set
\be
O_{\Lambda}(k)=\frac{1}{2}\expval{\hat{O}^{+}_{k}\hat{O}_{k}
+\hat{O}_{k}\hat{O}_{k}^{+}}_{\Lambda}
\end{equation}
From (\ref{ineqduhamel}) we get the two inequalities
\begin{align}
\abs{\mu}(\hat{B}_{k}^{+},\hat{B}_{k})_{\Lambda}^{1/2}&
\geq\abs{\nu}(\hat{C}_{k}^{+},\hat{C}_{k})_{\Lambda}^{1/2}-
\frac{1}{\sqrt{\beta}}\expval{[\hat{A}_{k}^{+},\hat{K}_{k}]}_{\Lambda}^{1/2}\\
\abs{\nu}(\hat{C}_{k}^{+},\hat{C}_{k})_{\Lambda}^{1/2}&
\geq\abs{\mu}(\hat{B}_{k}^{+},\hat{B}_{k})_{\Lambda}^{1/2}-
\frac{1}{\sqrt{\beta}}\expval{[\hat{A}_{k}^{+},\hat{K}_{k}]}_{\Lambda}^{1/2}
\end{align}
Using the upper and lower bounds (\ref{upperbound}) and (\ref{lowerbound}) we have
\begin{align}
\abs{\mu}B_{\Lambda}(k)^{1/2}&\geq
\abs{\nu}C_{\Lambda}(k)^{1/2}f(h_{\Lambda}(\hat{C}_{k}))^{1/2}-
\frac{1}{\sqrt{\beta}}\expval{[\hat{A}_{k}^{+},\hat{K}_{k}]}_{\Lambda}^{1/2}
\label{ineq3}\\
\abs{\nu}C_{\Lambda}(k)^{1/2}&\geq
\abs{\mu}B_{\Lambda}(k)^{1/2}f(h_{\Lambda}(\hat{B}_{k}))^{1/2}-
\frac{1}{\sqrt{\beta}}\expval{[\hat{A}_{k}^{+},\hat{K}_{k}]}_{\Lambda}^{1/2}
\label{ineq4}
\end{align}
From the two last inequalities and the behavior of $f(u)$ for $u\rightarrow 0$,
 it is easy to deduce the following

\vspace{5mm}

{\bf Theorem}

{\it Assume there exist three local observables $A_x$, $B_x$, $C_x$ satisfying
(\ref{commutation}). Suppose also that for a given $k$, there exist 
three positive constants $a_{k}$, $b_{k}$ and $c_{k}$
independent of $\abs{\Lambda}$  such that
$\expval{[\hat{A}_{k}^{+},[H_{\Lambda},\hat{A}_{k}]]}_{\Lambda}\leq a_{k}$,
$\expval{[\hat{B}_{k}^{+},[H_{\Lambda},\hat{B}_{k}]]}_{\Lambda}\leq b_{k}$ and
$\expval{[\hat{C}_{k}^{+},[H_{\Lambda},\hat{C}_{k}]]}_{\Lambda}\leq c_{k}$.
Then for any temperature $\beta^{-1}$
\be
\lim_{\Lambda\rightarrow\bbZ^d}\frac{1}{\abs{\Lambda}}B_{\Lambda}(k)\neq 0
\end{equation}
if and only if
\be
\lim_{\Lambda\rightarrow\bbZ^d}\frac{1}{\abs{\Lambda}}C_{\Lambda}(k)\neq 0
\end{equation}}

In the last theorem, it is important that the expectation values of the double
commutators are $O(1)$ with respect to the volume. In the Appendix, we show
that this is indeed the case for the Hamiltonian (\ref{hamiltonian}) and a
large class of observables. It turns out however that this assumption can often be checked explicitely in
concrete applications.

\section{Applications}

We wish to present an illustration of the theorem for
the dynamic Holstein model. This is a model for molecular crystals which retains
the interaction of itinerant electrons with breathing modes of the molecules.
These modes are described by Einstein oscillators of frequency $\omega$
attached to each site $x\in\Lambda$. Let $q_x$, $p_x$ be the position and
momentum of the oscillators (with $[q_x,p_y]=\ii\delta_{xy}$). The Hamiltonian
of the model is
\be
H_{\Lambda}=\sum_{x,y\in\Lambda}t_{xy}c_{x}^{+}
c_{y}+U\sum_{x\in\Lambda}q_{x}(n_{x}-\frac{1}{2})
+\frac{1}{2}\sum_{x\in\Lambda}(p_{x}^{2}+\omega^{2}q_{x}^{2})
\label{hamiltonian2}
\end{equation}
Choosing $A_{x}=p_{x}$ we obtain
\be
[H_{\Lambda},p_x]=\ii U(n_{x}-\frac{1}{2})+\ii\omega^{2}q_x
\label{commutation2}
\end{equation}
so that $B_{x}=(n_{x}-\frac{1}{2})$, $C_{x}=q_x$, $\mu=\ii U$ and
$\nu=\ii\omega^{2}$. To apply the theorem we
have to check that $\expval{[\hat{O}_{k}^{+},[H_{\Lambda},\hat{O}_{k}]]}_{\Lambda}=O(1)$
for $\hat{O}_{k}=\hat{A}_{k}$, $\hat{B}_{k}$ and $\hat{C}_{k}$. 
By explicit calculation we have
\begin{align}
[\hat{A}_{k}^{+},[H_{\Lambda},\hat{A}_{k}]]=&\omega^2\\
[\hat{B}_{k}^{+},[H_{\Lambda},\hat{B}_{k}]]=&\frac{1}{\abs{\Lambda}}
\sum_{x,y\in\Lambda}
2(\cos(k\cdot (x-y))-1)t_{xy}c_{x}^{+}c_{y}\label{B}\\
[\hat{C}_{k}^{+},[H_{\Lambda},\hat{C}_{k}]]=&1
\end{align}
The double commutators with $\hat{A}_{k}$ and $\hat{C}_{k}$ are constant 
and thus equal to their expectation values. For $\hat{B}_{k}$ we have
\be
\abs{\expval{[\hat{B}_{k}^{+},[H_{\Lambda},\hat{B}_{k}]]}}\leq
\frac{1}{\abs{\Lambda}}\sum_{x,y\in\Lambda}
\abs{2(\cos(k\cdot (x-y))-1)}\abs{t_{xy}}
\abs{\expval{c_{x}^{+}c_{y}}}_{\Lambda}\leq
\frac{4}{\abs{\Lambda}}\sum_{{x},{y}\in\Lambda}\abs{t_{xy}}=O(1)
\end{equation}
where we used that for fermionic operators $\abs{\expval{c_{x}^{+}c_{y}}}\leq
1$. Thus we can conclude that
$\lim_{\Lambda\rightarrow\bbZ^d}\frac{1}{\abs{\Lambda}}\expval{(\hat{n}_{-k}-\frac{1}{2})
(\hat{n}_{k}-\frac{1}{2})}_{\Lambda}\neq 0$ if and only if
$\lim_{\Lambda\rightarrow\bbZ^d}\frac{1}{\abs{\Lambda}}
\expval{\hat{q}_{-k}\hat{q}_{k}}_{\Lambda}\neq
0$.

It follows from this result that if a Peierls instability occurs at a given $k$
(i.e. a
distortion of the lattice with long-range order for $q_x$) there is at the same
time long-range order in the electronic density. The occurence of a Peierls
instability for $k=\pi$ at low temperature and half-filling
has in fact been proven for the static model
where the kinetic energy of the phonons in (\ref{hamiltonian2}) is absent
\cite{LeMa}. The
present analysis remains unchanged in the static case because even if the term
$p_{x}^2$ is absent from the Hamiltonian we still have the right to choose
$A_x=p_x$ so that (\ref{commutation2}) remains unchanged.

In one dimension, for the spinless case the decay of the fermionic correlation
functions for the ground state has been analyzed rigorously \cite{Benfatto} at
weak coupling $U$ by using renormalisation group techniques.
The results of \cite{Benfatto} imply that these correlations decay algebraically
and therefore we can conclude that no long-range order is present at the level of
$q_x$ as well.

An interesting application of our theorem concerns the comparison of critical
temperatures in
systems with several types of degrees of freedom and several order
parameters. An example is the Holstein model itself. Another example are the so-called
ferrofluids where one has magnetic (spin) degrees of freedom as well as
translational degrees of freedom (see for example \cite{gruber,zagrebnov}).
A relevant question is whether there exist two different critical
temperatures associated to each order parameters or only one critical
temperature where both order parameters simultaneously develop a non zero value.
For example in the ferrofluids, it is of interest to decide whether
liquid-vapor and ferro-paramagnetic phase transitions occur at the same
critical temperature or not.
Although our theorem is a statement about existence of long-range order rather
than about the order parameter, it suggests that when it is applicable there is
only one critical temperature. This fact had been proven earlier for the particular
case of the static Holstein model in \cite{LeMa}.

A related question that our basic result leaves open is whether the behavior 
of the two correlation functions at a critical point is governed by
two independent length scales. Our discussion above suggests that
 (when our result is applicable) this is
not the case and there is a single diverging length scale 
 involved in both correlations.

\vspace{5mm}

{\bf Acknowledgements}

We would like to thank C. Gruber, J. Ruiz and V. A. Zagrebnov for discussions.
The work of C.-A. Piguet was supported by 
the Swiss National Science Foundation.

\section*{Appendix}

In order to apply our theorem, it is necessary that all the expectation values
of the double commutators are $O(1)$. In this appendix, we check that this is
the case for the Hamiltonian (\ref{hamiltonian}) and a large class of
observables.
The double commutators we have to discuss are
\be
\expval{[\hat{O}_{k}^{+},[H_{\Lambda},\hat{O}_{k}]]}_{\Lambda}=
\frac{1}{\abs{\Lambda}}\sum_{z,z'\in\Lambda}\e{\ii k\cdot (z'-z)}
\expval{[O_{z'}^{+},[H_{\Lambda},O_{z}]]}_{\Lambda}
\label{a1}
\end{equation}
where $O_{z}$ is a local observable centered around $z$ of the form
\be
O_{z}=c_{z+\delta_{1}}^{+}\ldots c_{z+\delta_{N}}^{+}c_{z+\epsilon_{1}}\ldots
c_{z+\epsilon_{M}}
\label{a21}
\end{equation}
where $\delta_{i},\;i=1,\ldots,N$ and $\epsilon_{j},\;j=1,\ldots,M$ are fixed
vectors and $N$ and $M$ are fixed numbers. The subsequent discussion provides a proof that (35) is $O(1)$ when $N+M$ is even. If $N+M$ is odd we have only 
heuristic arguments presented at the end.

 We give the detailed calculation for
the kinetic term of (\ref{hamiltonian}) (for simplicity we omit spin indices).
The  interaction terms can be  treated in a similar way.
Let us evaluate the commutator of $O_{z}$ with the kinetic term
\begin{multline}
[\sum_{x,y\in\Lambda}t_{xy}c_{x}^{+}c_{y},O_{z}]=\sum_{i=1}^{N}
\sum_{x,y\in\Lambda}t_{xy}\delta_{y,z+\delta_{i}}
c_{z+\delta_{1}}^{+}\ldots c_{z+\delta_{i-1}}^{+}
c_{x}^{+}c_{z+\delta_{i+1}}^{+}\ldots
c_{z+\delta_{N}}^{+}c_{z+\epsilon_{1}}\cdots c_{z+\epsilon_{M}}\\
-\sum_{j=1}^{M}
\sum_{x,y\in\Lambda}t_{xy}\delta_{x,z+\epsilon_{j}}
c_{z+\delta_{1}}^{+}\ldots c_{z+\delta_{N}}^{+}
c_{z+\epsilon_{1}}\ldots c_{z+\epsilon_{j-1}}c_{y}c_{z+\epsilon_{j+1}}
\ldots c_{z+\epsilon_{M}}
\label{a2}
\end{multline}
The commutator of $O_{z'}^{+}$ with the first term of the
right-hand side of the last equation reads
\begin{multline}
\sum_{i=1}^{N}\sum_{x,y\in\Lambda}t_{xy}\delta_{y,z+\delta_{i}}
(c_{z'+\epsilon_{M}}^{+}\ldots c_{z'+\epsilon_{1}}^{+}c_{z'+\delta_{N}}\ldots
c_{z'+\delta_{1}}
c_{z+\delta_{1}}^{+}\ldots c_{z+\delta_{i-1}}^{+}\cdot\\
c_{x}^{+}c_{z+\delta_{i+1}}^{+}\ldots
c_{z+\delta_{N}}^{+}c_{z+\epsilon_{1}}\ldots c_{z+\epsilon_{M}}\\
-c_{z+\delta_{1}}^{+}\ldots c_{z+\delta_{i-1}}^{+}
c_{x}^{+}c_{z+\delta_{i+1}}^{+}\ldots
c_{z+\delta_{N}}^{+}c_{z+\epsilon_{1}}\ldots c_{z+\epsilon_{M}}
c_{z'+\epsilon_{M}}^{+}\ldots c_{z'+\epsilon_{1}}^{+}c_{z'+\delta_{N}}\ldots
c_{z'+\delta_{1}})
\label{a3}
\end{multline}

Using the fermionic commutation relations, we can drive in both terms the
creation operators to the left and the annihiliation operators to the right.
Doing this we obtain a finite number of
terms involving {\it two} Kronecker
symbols of the type $\delta_{y,z+\delta_{i}}\delta_{z'+\mu,z}$ or
$\delta_{y,z+\delta_{i}}\delta_{z'+\mu,x}$, with $\mu$ a linear combination
of $\delta$'s and $\epsilon$'s, plus a supplementary term
containing only {\it one} Kronecker symbol. The latter one is
\begin{multline}
(-1)^{N^2}(1-(-1)^{(M+N)^2})
\sum_{i=1}^{N}\sum_{x,y\in\Lambda}t_{xy}\delta_{y,z+\delta_{i}}
(c_{z'+\epsilon_{M}}^{+}\ldots c_{z'+\epsilon_{1}}^{+}
c_{z+\delta_{1}}^{+}\ldots c_{z+\delta_{i-1}}^{+}
c_{x}^{+}\cdot\\
c_{z+\delta_{i+1}}^{+}\ldots
c_{z+\delta_{N}}^{+}
c_{z'+\delta_{N}}\ldots c_{z'+\delta_{1}}
c_{z+\epsilon_{1}}\ldots c_{z+\epsilon_{M}})
\label{a4}
\end{multline}

We can now replace these expressions into (\ref{a1}). Let us first suppose that $M+N$
is even. In this case the term (\ref{a4}) vanishes and only the terms with
two Kronecker symbols contribute.
It follows from the
previous discussion that the contribution of (\ref{a3}) to (\ref{a1}) is
bounded by
\be
\frac{1}{\abs{\Lambda}}\sum_{i=1}^{N}\sum_{x,y,z,z'\in\Lambda}
\Delta_{i}(x,y,z,z')\abs{t_{xy}}\abs{\expval{\ldots c^{+}\ldots
c\ldots}_{\Lambda}}
\label{a5}
\end{equation}
where $\Delta_{i}(x,y,z,z')$ is one of the following: $\delta_{y,z+\delta_{i}}
\delta_{z'+\mu,z}$ or $\delta_{y,z+\delta_{i}}\delta_{z'+\mu,x}$.
The expectation value of the product of $c^{\#}$ can be bounded by 1 since the
norm of the fermionic operators is less than 1. Then using the integrability of
$t_{xy}$, the two Kronecker symbols and the factor $1/\abs{\Lambda}$,
 we conclude that (\ref{a5}) is bounded by some constant uniformly in
the volume.

It is  possible to perform a similar discussion for the second
term in the right-hand side of $(\ref{a2})$
to conclude that the double commutators
of local observables of the type (\ref{a21}) with $H$ have to be $O(1)$ with respect to
 the volume in the case $N+M$ even.

In the case where $N+M$ is odd, the discussion for the terms with two
Kroneckers symbols is the same as in the case of $N+M$ even. However, there
remains the term (\ref{a4}). Replacing it in (\ref{a1}) produces a term bounded
by
\begin{multline}
\frac{1}{\abs{\Lambda}}\sum_{i=1}^{N}\sum_{x,y,z,z'\in\Lambda}
\delta_{y,z+\delta_{i}}\abs{t_{xy}}\abs{\expval{c_{z'+\epsilon_{M}}^{+}\ldots
c_{z'+\epsilon_{1}}^{+}
c_{z+\delta_{1}}^{+}\ldots c_{z+\delta_{i-1}}^{+}
c_{x}^{+}c_{z+\delta_{i+1}}^{+}\ldots
c_{z+\delta_{N}}^{+}\\
c_{z'+\delta_{N}}\ldots c_{z'+\delta_{1}}
c_{z+\epsilon_{1}}\ldots c_{z+\epsilon_{M}}}
_{\Lambda}}
\label{a6}
\end{multline}
In this last expression, the averages of the product of creation and
annihiliation operators are matrix elements of the reduced density matrix
$\rho_{M+N}$. In \cite{yang} Yang conjectured that long range  correlations
(ODLRO) may appear in such matrix elements only
for $N+M$ {\it even}. We are not aware of a general proof of this conjecture but it suggests that the matrix element in (41) decays as 
 $\abs{x-z'}\rightarrow\infty$. Assuming the decay is quick enough the sum
 over $z^\prime$ in (41) converges, which then leads to an upper bound
\be
\frac{c}{\abs{\Lambda}}\sum_{i=1}^{N}\sum_{x,y,z\in\Lambda}
\delta_{y,z+\delta_{i}}\abs{t_{xy}}
\label{a7}
\end{equation}
where $c$ is a constant independent of the volume.
Summing over $z$, using the integrability of the hopping amplitudes and the
term $1/\abs{\Lambda}$, we have that (\ref{a7}), and thus
(\ref{a6}), are bounded by a constant
uniformly in the volume.

\end{document}